\documentclass[onefignum,final]{siamltex}
\usepackage{epsfig}

\def\bivec#1{\vbox{\ialign{##\crcr $\leftrightarrow$\crcr\noalign{
 \kern-1pt \nointerlineskip}$\hfil\displaystyle{#1}\hfil$\crcr}}}

\title{Multiscale coupling of molecular dynamics and hydrodynamics: 
application to DNA translocation through a nanopore
}

\author{Maria G. Fyta\footnotemark[2] \footnotemark[3]\and
Simone Melchionna\footnotemark[4]\and
Efthimios Kaxiras\footnotemark[2] \footnotemark[3] \and
Sauro Succi\footnotemark[5]}

\begin{document}

\maketitle
\renewcommand{\thefootnote}{\fnsymbol{footnote}}

\footnotetext[2]{Department of Physics and Division of Engineering and Applied
Sciences, Harvard University, Cambridge, MA, USA.}
\footnotetext[4] {INFM-SOFT, Department of Physics,
 Universit\`a di Roma {\it La Sapienza}, P.le A. Moro 2, 00185 Rome, Italy.}
\footnotetext[5]{Istituto Applicazioni Calcolo, CNR, 
Viale del Policlinico 137, 00161, Roma, Italy.}
\footnotetext[3] {This work was supported primarily by the Nanoscale Science
and Engineering Center, funded by the National Science Foundation,
Award Number PHY-0117795.}

\markboth{Fyta, Melchionna, Kaxiras, \& Succi}{Multiscale coupling of molecular dynamics and hydrodynamics}

\begin{abstract}
We present a multiscale approach to the modeling of 
polymer dynamics in the presence of a fluid solvent.
The approach combines Langevin Molecular Dynamics (MD)
techniques with a mesoscopic 
Lattice-Boltzmann (LB) method for the solvent dynamics. 
A unique feature of the present approach is that hydrodynamic 
interactions between the solute macromolecule and the aqueous solvent are
handled explicitly, and yet in a computationally 
tractable way due to the dual particle-field nature of the LB solver.
The suitability of the present LB-MD multiscale approach
is demonstrated for the problem of polymer fast translocation through
a nanopore.
We also provide an interpretation of our results in the context 
of DNA translocation through a nanopore, a problem that  
has attracted much theoretical and experimental attention recently.
\end{abstract}



\pagestyle{myheadings}
\thispagestyle{plain}

\section{Introduction}

Mathematical modeling and computer simulation of biological
systems is in a stage of burgeoning growth.
Advances in computer technology but also, 
perhaps more importantly, breakthroughs
in simulational methods are helping to reduce the gap between
quantitative models and actual biological behavior.
The main challenge remains the wide and disparate range
of spatio-temporal scales involved in the dynamical evolution of
complex biological systems.  In response to this challenge, 
various strategies have been developed recently, 
which are in general referred to as ``multiscale modeling''. 
Some representative examples include hybrid continuum-molecular dynamics
algorithms \cite{HCMD}, heterogeneous multiscale methods \cite{HMM},
and the so-called equation-free approach \cite{EFREE}.
These methods combine different levels
of the statistical description of matter (for instance, continuum and 
atomistic) into a composite computational scheme, in which information is 
exchanged through appropriate hand-shaking regions between the scales.
Vital to the success of this information exchange procedure
is a careful design of proper hand-shaking interfaces.

Kinetic theory lies naturally between the continuum and atomistic 
descriptions, and should therefore 
provide an ideal framework for the development of robust 
multiscale methodologies.
However, until recently, this approach has been hindered
by the fact that the central equation of kinetic theory, that is, 
the Boltzmann equation, was perceived as 
an equally demanding approach as molecular dynamics
from the computational point of view, and of very limited
use for dense fluids due to the lack of many-body correlations.
As a result, multiscale modeling of nanoflows
has developed mostly in the direction of the continuum/molecular 
dynamics paradigm \cite{HCMD}. 

Over the last decade and a half, major developments in lattice kinetic 
theory \cite{LBE,LBEORIG} are changing the scene.
Minimal forms of the Boltzmann equation
can be designed on the lattice, which quantitatively describe the behavior
of fluid flows in a way that is often computationally more advantageous
than the continuum approach based on the Navier-Stokes equations.
Moreover, lattice kinetic theory has proven capable of dealing with
complex flows, such as flows with phase transitions and strong heterogeneities,
for which continuum equations are exceedingly difficult to solve, if
at all known (for a recent review see \cite{LGA05}).
These advances have opened the road to developing 
new mesoscopic multiscale solvers \cite{MULBE}.
The present work provides a successful implementation of such an approach.
We will focus on the coupling of a {\it mesoscopic} fluid
solver, the lattice Boltzmann method, with simulations at the atomistic scale
employing explicit molecular dynamics.
A unique feature of our approach 
is the dual nature of the mesoscopic kinetic solver, which propagates 
coarse-grained information (the single-particle
Boltzmann probability distribution), along straight particle trajectories.
This dual field/particle nature greatly facilitates the coupling between
the mesoscopic fluid and the atomistic levels, both on conceptual and 
computational grounds.

The paper is organized as follows. In Section $\S$\ref{basic} we present the
basic elements of the multiscale methodology, namely
the Lattice Boltzmann treatment of the fluid solvent, and its coupling to
a Molecular Dynamics simulation of the solute biopolymer.
In Section $\S$\ref{applyMethod}, we present an application of this multiscale
methodology to the problem of long polymer translocation through a 
nanopore; 
in particular, we analyze in detail the role of hydrodynamics 
in accelerating the translocation process.
In Section $\S$\ref{relate2dna} we elaborate on the relevance of our results 
to the problem of DNA translocation, which has attracted much 
theoretical and experimental attention recently.
We conclude in Section $\S$\ref{outlook} with general remarks and outlook for
 future extensions.

\section{Lattice-Boltzmann - Molecular-Dynamics multiscale methodology}
\label{basic}

We consider the generic problem of tracing the dynamic evolution
of a polymer molecule interacting with a fluid solvent.
This involves the simultaneous interaction of several
physical mechanisms, often acting on widely separate temporal 
and spatial scales. Essentially, these interactions can be classified 
in three distinct categories as
solute-solute, solvent-solvent and solvent-solute.
The first category includes the conservative many-body interactions among 
the single monomers in the polymer chain. 
Being atomistic in nature, these interactions
usually set the shortest scale in the overall multiscale process.
They are typically handled by Molecular Dynamics techniques for
constrained molecules.
The second category, the solvent-solvent interactions,
refer to the dynamics of the solvent molecules, 
which are usually dealt with by a continuum fluid-mechanics
approach; in the present work
these will be described by the mesoscopic Lattice Boltzmann equation.  
The second and third category have also been handled by simulating 
the solvent explicitly via molecular dynamics, implicit solvent particles 
via Brownian dynamics including hydrodynamic interactions, 
or solving the corresponding Fokker-Planck equation \cite{kroger}. Finally, 
the solvent-solute dynamics will be treated by augmenting the molecular
dynamics side with dissipative fluid-molecule interactions (Langevin picture)
and including the corresponding reaction terms in the fluid-kinetic equations. 

\subsection{Atomistic dynamics}

We consider a polymer consisting of $N$ monomer
units (also referred to as beads).
The polymer is advanced in time according to the following
set of Molecular Dynamics-Langevin equations for the bead positions 
$\vec{r}_p$ and velocities $\vec{v}_p$:
\begin{eqnarray}
\label{MD}
\frac{d \vec{r}_p}{dt} &=& \vec{v}_p\\
m \frac{d \vec{v}_p}{dt} &=&
\vec{F_p}^c + \vec{F_p}^f + \vec{F_p}^r + \vec{F_p}^{\kappa},\; \,\,\,
p=1,N
\end{eqnarray}
where we distinguish four types of forces:
\begin{eqnarray}
\label{MDFORCES}
\vec{F_p}^{c~} &=&-\sum_q \partial_{\vec{r}_p} V(\vec{r}_p-\vec{r}_q)\\
\vec{F_p}^{f~} &=& \gamma (\vec{u}_p-\vec{v}_p)\\
\vec{F_p}^{r~} &=& m \vec{\xi}_p\\
\vec{F_p}^{\kappa} &=&-\lambda_p \partial_{\vec{r}_p} \kappa_p
\end{eqnarray}
The first term represents the conservative bead-bead 
interactions through a potential 
which we will take to have the standard $6-12$ Lennard-Jones form, 
\begin{equation}
V_{LJ}(r) = 4 \epsilon [(\sigma/r)^{12}-(\sigma/r)^6]
\label{LJ_potential}
\end{equation} 
truncated at a distance of r=$2^{1/6}\sigma$ \cite{WCA}. This
was combined with a harmonic part to account for the energy cost 
of distorting the angular degrees of freedom,
\begin{equation}
V_{ang}(\phi) =\frac{\kappa \phi^2}{2}
\label{Ang_potential}
\end{equation} 
with $\phi$ the relative angle between two consecutive bonds. 
Torsional motions will not be included in the present model, but 
can easily be incorporated if needed. 

We consider next the solute-solvent interactions.
The second term on the right-hand-side of Eq.(\ref{MD}) represents the 
mechanical friction between the single bead and the 
surrounding fluid, $\vec{v}_p$ being the bead velocity
and $\vec{u}_p$ the fluid velocity evaluated at the bead position.
In addition to mechanical drag, the polymer feels the effects of stochastic
fluctuations of the fluid environment, through the random 
term, $\vec{\xi}_p$, a Gaussian noise obeying the
fluctuation-dissipation relations: 
\begin{eqnarray*}
<\vec{\xi}_p>&=&0\\ 
<\vec{\xi}(\vec{r}_p,t) \vec{\xi}(\vec{r}_q,t')>&=&
\gamma (k_BT/m) V \delta(\vec{r}_p-\vec{r}_q) \delta(t-t')
\end{eqnarray*}
where $V$ is the volume of the cell to which beads $p$ and $q$ belong.
Finally, $\lambda_p \partial_{\vec{r}_p} \kappa_p$ is the reaction 
force resulting from $N-1$ holonomic constraints for molecules modelled 
with rigid covalent bonds:
\begin{eqnarray}
\label{kappa}
\kappa_p \equiv |\vec{r}_{p+1}-\vec{r}_p|^2 - r_0^2 = 0
\end{eqnarray}
$r_0$ being the prescribed bond length, and $\lbrace \lambda_p \rbrace$ is the
set of $N-1$ Lagrange multipliers conjugated to each constraint.
The usage of constraints instead of flexible bond lengths makes it possible to 
eliminate unimportant high-frequency intra-molecular motion which would render
the underlying LB propagation prone to numerical instabilities. 
In this way, the time-step of the Molecular Dynamics part 
can be increased by about one order of magnitude, as much as the overall 
efficiency of the LBMD method, as we shall discuss in section 
$\S$\ref{eff}.
Finally, in order to avoid spurious dissipation,
the bead velocities are required to
be strictly orthogonal to the relative displacements.
Given the second order atomistic dynamics, the velocities must obey the independent constraints:
\begin{eqnarray}
\label{kappadot}
\frac{d\kappa_p}{dt} = (\vec{r}_{p+1}-\vec{r}_p)\cdot(\vec{v}_{p+1}-\vec{v}_p) = 0
\end{eqnarray}
The constraints (\ref{kappa}), (\ref{kappadot}) are enforced over 
positions and momenta separately via the SHAKE \cite{SHAKE} and the 
RATTLE algorithms \cite{RATTLE}.
The implementation of these constraints requires the iterative solution of the
system of equations (\ref{kappa})-(\ref{kappadot}), typically accomplished
via standard Newton-Raphson techniques.

\begin{figure}
\begin{center}
\includegraphics[width=0.45\textwidth,angle=0]{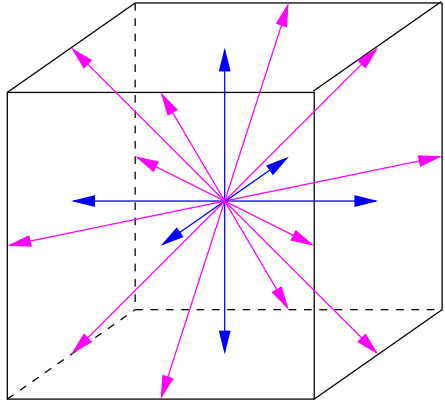}
\caption{\label{FIG1}The set of discrete speeds in the standard 19-speed 
3D lattice for the Lattice-Boltzmann method.}
\end{center}
\end{figure}

\subsection{Fluctuating Lattice Boltzmann method}

The Lattice Boltzmann equation is a minimal form of the Boltzmann
kinetic equation in which all details of molecular motion are
removed except those that are strictly needed to recover
hydrodynamic behavior at the macroscopic scale (mass-momentum
and energy conservation).
The result is an elegant equation for
the discrete distribution function $f_i(\vec{x},t)$
describing the probability to find a particle at lattice
site $\vec{x}$ at time $t$ with speed $\vec{v}=\vec{c}_i$.
More specifically, since we are dealing with nanoscopic flows, 
in this work we shall consider the
fluctuating Lattice Boltzmann equation
which takes the following form:
\begin{equation} 
\label{LBE}
f_i(\vec{x}+ \vec{c}_i \Delta t,t+\Delta t) = f_i(\vec{x},t) - 
\omega \Delta t (f_i-f_i^{eq})(\vec{x},t) +  F_i \Delta t + S_i \Delta t
\label{lbe2}
\end{equation} 
where $f_i(\vec{x},t)$ represents the probability of finding a fluid
particle at spatial location $\vec{x}$ and time $t$
with discrete speed $\vec{c}_i$.
The particles can only move along the links of a regular lattice 
defined by the discrete speeds, so that the 
synchronous particle displacements 
$\Delta \vec{x}_i = \vec{c}_i \Delta t$ never 
take the fluid particles away from the lattice.
For the present study, the standard three-dimensional 
19-speed lattice is used \cite{LBE} (see Figure \ref{FIG1}).
The right hand side represents the effect of intermolecular 
solvent-solvent collisions, through
a relaxation toward local equilibrium, $f_i^{eq}$, typically a 
second order (low-Mach) expansion in the fluid
velocity of a local Maxwellian with speed $\vec{u}$:
\begin{equation}
f_i^{eq} = w_i \rho \lbrace
1 + 
\beta \vec{u} \cdot \vec{c}_i +
\frac{\beta^2}{2}  [\vec{u} \vec{u} : 
(\vec{c}_i \vec{c}_i - \beta^{-1}{\rm \bivec{I}})]
\rbrace
\end{equation}
where $\beta=m_f/k_B T_f$ is the inverse fluid temperature (with $k_B$
the Boltzmann constant),
$w_i$ a set of weights normalized to unity, and {\rm \bf I} is 
the unit tensor in configuration space.
The relaxation frequency $\omega$ controls the fluid 
kinematic viscosity $\nu$ through the relation:
\begin{equation}
\nu= c_s^2 (1/\omega-\Delta t/2) 
\end{equation}
where $c_s $ is the sound speed in the solvent \cite{MULBE}.
Knowledge of the discrete distributions $f_i$ allows the calculation of the 
local density $\rho$, flow speed $\rho \vec{u}$ 
and momentum-flux tensor $\bivec{P}$, by 
a direct summation upon all discrete distributions:
\begin{eqnarray}
\rho(\vec{x},t)&=&\sum_{i} f_{i}(\vec{x},t)  \label{dens} \\
\rho \vec{u} (\vec{x},t)&=&\sum_{i} f_i(\vec{x},t) \vec{c}_i \label{vel} \\
\bivec{P} (\vec{x},t)&=&\sum_{i} f_i(\vec{x},t) \vec{c}_i \vec{c}_i 
\label{pre}
\end{eqnarray}
The diagonal component of the momentum-flux tensor gives
the fluid pressure, while the off-diagonal terms give the shear-stress.
Unlike in hydrodynamics, both quantities are available locally and at any
point in the simulation.

Thermal fluctuations are included through the source term $F_i$ which 
reads as follows (index notation)
\begin{eqnarray}
F_i = w_i\rho\{ F^{(2)}_{ab}  (c_{ia} c_{ib} - \beta^{-1} \delta_{ab}) + F^{(3)}_{abc} g_{iabc} \}
\label{fluct}
\end{eqnarray}
where $F^{(2)}$ is the fluctuating stress tensor (a $3\times 3$ stochastic matrix).
Consistency with the fluctuation-dissipation 
theorem at {\em all} scales requires the following
conditions
\begin{eqnarray}
\langle F^{(2)}_{ab}(\vec{x},t) F^{(2)}_{cd}(\vec{x}',t') \rangle = 
{\gamma k_B T \over m} \Delta_{abcd} \delta(\vec{x}-\vec{x}')\delta(t-t')
\label{fdt}
\end{eqnarray}
where $\Delta_{abcd}$ is the fourth-order Kronecker symbol \cite{ladd}.
$F^{(3)}$ is related to the fluctuating heat flux and 
$g_{iabc}$ is the corresponding basis in kinetic space, 
essentially a third-order
Hermite polynomial (full details are given in \cite{adhikari}).

The polymer-fluid back reaction is described through the source term
$S_i$, which represents the momentum
input per unit time due to the reaction of the polymer on 
the fluid population $f_i$:
\begin{equation}
S_i (\vec{x},t) 
= w_i  \beta
\sum_{p \in D(x)} [ \vec{F}_p^f + \vec{F}_p^r ] \cdot \vec{c}_i
\end{equation}
where $D(x)$ denotes the mesh cell to which the {\it p}$^{th}$ bead
 belongs.
The quantities on the left hand side in the above expression have to reside
on the lattice nodes, which means that the frictional and random
forces need to be extrapolated from the particle to the grid location. 

The use of a LB solver for the fluid solvent is 
particularly well suited to this problem because of the following reasons: \\
i) Free-streaming proceeds along straight trajectories.
   This is in stark contrast with hydrodynamics, in which
   fluid momentum is transported by its own space-time varying
   velocity field. Besides securing exact conservation of mass
   and momentum of the numerical scheme, this also
   greatly facilitates the imposition of geometrically
   complex boundary conditions.
\\
ii) The pressure field is available locally, with no 
    need of solving any (computationally expensive) Poisson
    problem for the pressure, like in standard hydrodynamics.
\\
iii) Unlike hydrodynamics, diffusivity is not represented by a 
     second-order differential operator, but
     it emerges instead from the first-order LB relaxation-propagation
     dynamics. The result is that the kinetic scheme can march 
     in time-steps which
     scale linearly, rather than quadratically, with the mesh resolution.
     This facilitates high-resolution down-coupling to atomistic scales.
\\
iv) Solute-solute interactions preserve their local nature since
    they are explicitly mediated by the solvent molecules through
    direct solvent-solute interactions.
    As a result, the computational cost of hydrodynamic interactions
    scales only linearly with the length of the polymer 
    (no long-range interactions).
\\
v) Since {\it all} interactions are local, the LB scheme 
    is ideally suited to parallel computing.

It is worth mentioning that more advanced Lattice Boltzmann models 
\cite{karlin1,karlin2}
could equally well be coupled to the atomistic dynamics.

\subsection{Time exchange}

The Molecular-Langevin-Dynamics solver is marched in time with a
stochastic integrator (due to extra non-conservative
and random terms), proceeding at a fraction of the LB time-step, 
\[
dt = \Delta t/M 
\]
The time-step ratio $M>1$ controls the scale separation between the
solvent and solute timescales.

The numerical solution of the stochastic equations is performed by means
of a modified version of the Langevin Impulse propagation scheme, 
derived from the assumption that the systematic forces are constant between 
consecutive time steps \cite{LIP}.
The propagation of the unconstrained dynamics proceeds according to
the scheme \cite{melchtoappear}
\begin{eqnarray}
~~~~~~~~~\tilde r_p&=& r_p(t) + \frac{dt}{2} v_p(t) \nonumber \\
~~~~~~~~~v_p^\star(t+dt)&=& e^{-\gamma dt} \left\{
v_p(t)+\left(\frac{e^{\gamma dt}-1}{\gamma m}\right) 
\left ( F(\tilde r_p)  + \gamma u_p\right) 
+
\left( e^{\gamma dt/2}+1 \right) C(dt) \right\} \nonumber \\
~~~~~~~~~r_p^\star(t+dt)&=& \tilde r_p + \frac{dt}{2} v_p(t+dt) 
\label{LImod}
\end{eqnarray}
where $C(dt)$ is an array of $3N$ gaussian random variables with
zero mean and variance $\frac{k_B T}{m}(e^{2\gamma dt} - 1)$, and 
$\{\tilde r\}$ represent temporary positions.
The propagator (\ref{LImod}) is particularly suitable for our purposes 
since it is second order accurate in time and robust, that is, it 
reduces to the symplectic Verlet algorithm for $\gamma \rightarrow 0$.
Moreover, at variance with the original Langevin Impulse scheme, the 
modified propagator allows for an unambiguous definition of velocities, 
which are needed to couple the
polymer to the hydrodynamic field of the surrounding solvent.
The particle positions and velocities corrected via the SHAKE and 
RATTLE algorithms read
$\vec{r}^\star_p(t+dt) \rightarrow \vec{r}_p(t+dt) $ and
$\vec{v}^\star_p(t+dt) \rightarrow \vec{v}_p(t+dt) $.
For consistency, in considering the momentum exchange with the solvent
the corrected velocities appear in the friction forces.
The MD cycle is repeated $M$ times, with the hydrodynamic field frozen
at time $t_n= n \Delta t$.

\subsection{Spatial exchange}

\begin{figure}
\begin{center}
\includegraphics[width=0.8\textwidth]{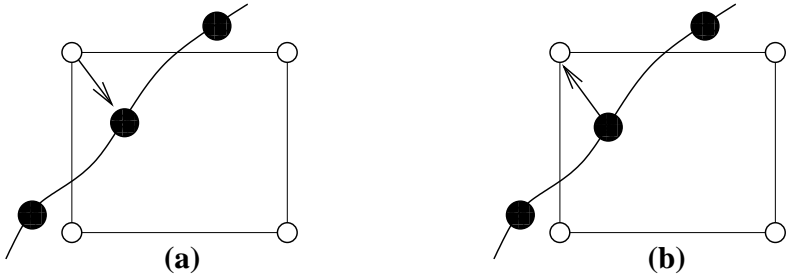}
\caption{\label{FIG2}Transfer of spatial information (a) from grid to particle, and (b) from particle to grid. Black spheres denote beads, while in white are the lattice sites.} 
\end{center}
\end{figure}

The transfer of spatial information from/to grid to/from particle locations 
is performed at each LB time-stamp $t_n=n \Delta t$.
To this purpose,on account of its simplicity, a simple nearest 
grid point (NGP) interpolation scheme is used (see Fig.\ref{FIG2}). 
Momentum conservation was checked to hold up to six digits.
With reference to a time slice $t_n=n \Delta t$, the 
pseudo-algorithm performing a single LB time-step, reads as follows
{\em
\begin{enumerate}
\item Interpolation of the velocity: $\vec{u}(\vec{x}) \rightarrow \vec{u}_p$ 
\item For $m=1,M$:
\begin{itemize}
\item[] Advance the molecular state from $t$ to $t+dt$ 
\end{itemize}
\item Extrapolation of the forces: $\vec{F_p} \rightarrow \vec{F}(\vec{x})$ 
\item Advance the Boltzmann populations from $t$ to $t+\Delta t$ 
\end{enumerate}
}
This time-marching can be formally represented by an operator-splitting
multi-step time procedure for two coupled 
kinetic equations describing the dynamic evolution 
for the fluid and the polymer distribution functions,
respectively \cite{BERNE}.
It is worth emphasizing that, while LB and MD with Langevin dynamics 
have been coupled before, notably in the study of single-polymer 
dynamics \cite{DUN}, to the best of our knowledge, this is
the first time such that coupling is put in place for {\it long} molecules 
of biological interest.

\subsection{Efficiency considerations}
\label{eff}

The total cost of the computation scales roughly like
\begin{equation}
t \sim (t_{LB} V  + t_{MD} M N) N_{LB}
\label{time}
\end{equation}
where $t_{LB}$ is the CPU time required to update a single LB site per
timestep and $t_{MD}$ is the CPU time to update a single bead per timestep,
$V$ is the volume of the computational domain
in lattice units and $N$ is the number of polymer beads, with M
the LB-MD time-step ratio.
Finally, $N_{LB}$ is the number of LB timesteps.
In the above equation, $t_{MD}$ includes the overhead of LB-MD coupling.
Note that $t_{MD}$ is largely independent of $N$ because i) the 
LB-MD coupling is local, ii) the forces are short ranged and 
iii) the SHAKE/RATTLE algorithms are empirically known 
to scale linearly with the number of constraints. 

Regarding the cost of the LB section, this is known to scale 
linearly with the volume occupied by the solvent.
For the case where polymer concentration is kept constant, the volume needed
to accommodate a polymer of $N$ beads should
scale approximately as $N^{1.8}$; however, for translocation studies 
such as those discussed later in this paper, 
we shall consider a box of given volume, independently on the polymer length.

From the above expression it is clear that $M$ should be chosen as small
as possible, consistent with the requirement of providing a realistic
description of the polymer dynamics.
In the present simulation we typically choose $M$ between $5$ and $20$,
depending on the parameters of the simulation, particularly the 
temperature. 
This means that we are taking the LB representation close to the molecular
scale. We will return to this important issue in the 
quantitative discussion of the physical application.

A tentative estimate of the computational cost proceeds as follows:
Assuming $250$ flops/site/LB-step and $2500$ flops/bead/MD-step (including
the LB-MD coupling overhead),
and an effective processing speed of $100$ Mflop/s, the evolution
over $30,000$ LB steps=$150,000$ MD steps of a typical $80\times40\times40$ grid 
and $400$ beads set-up, would take about:
\[
t=30,000\times\big[250\times(80\times40\times4)+2500\times5\times400\big]/10^8=(9600+1500) sec \sim 3 hrs,
\]
which is in reasonable agreement with the simulation time 
observed with the present version of the code ($7 hrs$), including 
the relative MD/LB cost ($\simeq 1:4$). 

We wish to emphasize that the key feature of the LB-MD
approach, namely {\it linear} scaling of the CPU cost with the number
of beads (at constant volume) is indeed observed.
In fact, the execution times for $50$, $100$ and $400$ beads are 
$0.433$, $0.489$, and $0.882$ sec/step, 
respectively on a 2GHz AMD Opteron processor. By excluding 
hydrodynamics, these numbers become $0.039$, $0.075$, and $0.318$ sec/step.
It is worth mentioning that thus far, no effort has been directed to
code optimization; it is
quite possible that careful optimization may lower the execution time
by an order of magnitude.

\subsection{Validation tests}

The static and dynamic behavior of the DNA chain obtained by our methodology
has been compared to the scaling predictions for a single chain at
infinite dilution. Given the structure factor 
$S_f(k)=\frac{1}{N}\sum_{i,j}\langle e^{ik\cdot(r_{i}-r_{j})}\rangle$,
standard theory predicts the scaling law $S_f(k)=Ng(kR_{g})$, where
$g(y)$ is a universal function and 
$R_{g}=(1/2N^{2})\sum_{i,j}\langle(r_{i}-r_{j})^{2}\rangle$
is the gyration radius. For large $k$, the structure factor is
independent of $N$, and it follows\[
S_f(k)\propto k^{-1/\mu}\]
where experiments, theory and simulations agree on the scaling exponent
value $\mu\simeq0.584$. The static scaling law is not affected by
the presence of hydrodynamics. However, verification of the scaling
law and attainment of the scaling regime for large enough chains is
a good check for the correctness of our simulation scheme and for
the subsequent validation of the hydrodynamic behavior. 

The dynamic behavior of the chain is deeply affected by the presence of hydrodynamic
interactions. The standard picture of polymer dynamics is based on
the Rouse (no hydrodynamics) or Zimm (hydrodynamic) description in
terms of an underlying gaussian chain. In this case, the chain intermediate
scattering function $I_S(k,t)=\frac{1}{NS_f(k)}\sum_{i,j}\langle e^{ik\cdot[r_{i}(t)-r_{j}(0)]}\rangle$
should follow the universal behavior\[
I_S(k,t)=\tilde{g}(D_fR_{g}^{x-2}tk^{x})\]
where $\tilde{g}(y)$ is another universal function and $D_f$ is
the center of mass diffusion constant. Having introduced the dynamic
scaling exponent $\mu'$ via $D_f\sim N^{\mu'}$, the exponent $x$
is found to be $x=2+\mu'/\mu$. According to Zimm theory, $\mu'=\mu$
and $x=3$ \cite{DUN} while, according to Molecular Dynamics simulations,
it appears that the actual value is somehow lower, i.e. $x\simeq2.9$
\cite{valid1}.

\begin{figure}
\begin{center}
\includegraphics[width=1.0\textwidth]{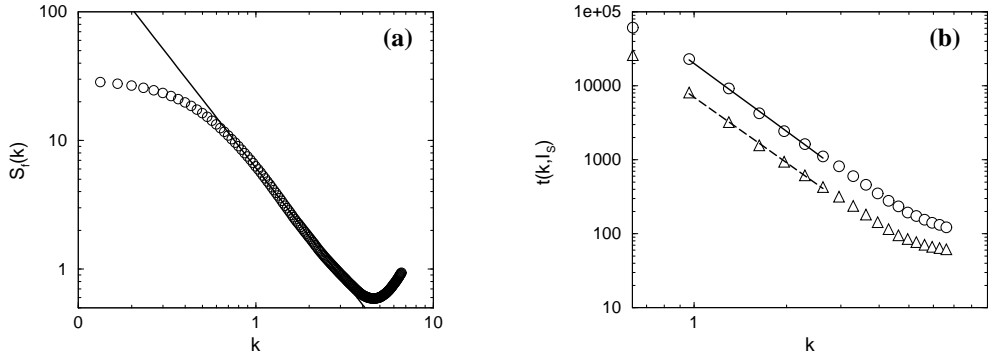}
\caption{\label{FIG3}(a) Log-log plot of the structure factor of a polymer 
in solution made of 30 monomers. The straight line is the power law fit in 
the range $1<k<3$ with exponent $\mu=0.58$. (b) Log-log parametric plot of 
$t(I_S,k)$ vs $k$ for $I_S=0.3$ (circles) and $I_S=0.6$ (triangles). 
The lines represent the power law fits within the scaling region 
$1<k<3.0$ with exponent $x=2.9$.}
\end{center}
\end{figure}

We have considered a chain made of 30 monomers with bead-bead Lennard-Jones
parameters taken from previous studies of chains simulated via Brownian
dynamics \cite{valid2} 
($\sigma=0.65$, $\epsilon/k_BT=1.0$, bond length $r_0=0.945$)
in a simulation box of edge $60$. This choice was motivated to verify
the range of scaling behavior as compared to previous numerical results.
We have computed the structure factor, as reported in Fig.\ref{FIG3}(a), and
observed that the scaling regime is clearly visible for $1<k<3$ with
an exponent equal $\mu=0.58\pm0.1$. In the range of $k$ vectors
where the static scaling holds, the dynamic scaling has been checked
by considering, for given values of the computed scattering function,
the loci of points $t(k,I_S)=\tilde{g}^{-1}(I_S)k^{-x}$ and by fitting
via a power law curve (see ref. \cite{valid1} for details). As illustrated
in Fig.\ref{FIG3}(b), the resulting scaling exponent is found to 
be $x=2.9\pm0.1$, in excellent agreement with the expected value 
and similar to previous simulation results on single polymers 
surrounded by a Lattice Boltzmann fluid \cite{valid2}. 
Moreover, by applying a heuristic argument \cite{valid1}, we
have verified that within the scaling regime, the finite size of the
periodic box was not biasing the data. 

\section{Application: polymer translocation through nanopore}
\label{applyMethod}

The scheme described above is general and applicable to any 
situation where a long polymer is moving in a solvent.  This 
motion is of great interest for a fundamental
understanding of polymer dynamics in the presence of the solvent.
For example, the translocation of a polymer through a pore of very small size
(of order the separation between monomers), is a process in which the coupling 
of the molecular motion to the solvent dynamics may be of crucial 
significance.  In this section, 
we will therefore provide a detailed discussion of the polymer dynamics
in the presence of a solvent for the example of translocation 
through a nanopore but without reference to a specific physical system.
In the next section we explore the relevance of these results to 
DNA translocation through a nanopore.

\subsection{Initial and Boundary conditions}

The polymer is initialized via a standard self-avoiding
random walk algorithm and further relaxed to equilibrium by standard 
Molecular Dynamics.
The solvent is initialized with the equilibrium distribution
corresponding to a constant density $\rho_0$ and zero
macroscopic speed $\vec{u}=0$.

Boundary conditions for the fluid are periodic at 
inlet/outlet sections, and zero-speed at rigid walls, using
the standard bounce-back rule \cite{LBE}. 
For the polymer, periodicity is again imposed at inlet/outlet, whereas
the interaction with rigid walls is handled by a Lennard-Jones
potential with specific wall-polymer parameters $\sigma_{wall}=1.5$
and $\epsilon_{wall}=10^{-3}$ in LB units.
The connection between slip-flow at the wall and intermolecular
solid-fluid interactions shall be the objects of future research.
 
\subsection{Numerical set-up}

We consider a three-dimensional box of size
$N_x h \times N_y h \times N_z h$ 
lattice units, with $h$ the spacing between lattice points.
We will take $N_x = 2 N_y$, $N_y = N_z$; the separating wall is located in 
the mid-section of the $x$ direction, at $x=h N_x/2$
with $N_x = 80$.
At $t=0$ the polymer resides entirely in the right chamber 
at $x>h N_x/2$. At the center of the separating wall, a square
hole of side $d_{hole}=2h$ is opened, through which the
polymer can translocate from one chamber to the other.
Translocation is induced by a constant electric field which
acts along the $x$ direction, and is confined in a rectangular
channel of size $2h \times h \times h$ along the streamline 
($x$ direction) and cross-flow ($y,z$ directions). 
The spatial coarse-graining is such that the presence of the solvent
as well as electrostatic forces acting due to charges on the polymer
are neglected altogether as being of secondary importance 
compared to hydrodynamics. 

Here and throughout we work in lattice Boltzmann units, in which 
length and time are measured in units of the lattice spacing
$h = \Delta x$ and time-step $\Delta t$, respectively.
Mass is defined as $m = m_{LB} m_{sol}$. The dimensionless mass 
$m_{LB}$ used in the simulations is set to unity,
which means that mass is measured in units of the solvent mass $m_{sol}$.
This choice is not restrictive since the present approach is used to
model incompressible flows in which density is a parameter 
which can be rescaled by any arbitrary factor.
However, it is of some interest to estimate the number of solvent molecules
represented by a single LB computational molecule, since the inverse of
this number conveys a measure of the importance of statistical
fluctuations at the scale of the lattice spacing $\Delta x$.
Let $S_N$ be this number, which will be defined as 
$\frac{\rho_{sol} h^3}{\rho_{LB} m_{sol}}$,
where $\rho_{LB}$ is the dimensionless density used in the LB simulations.
In order for the Boltzmann probability distribution to make sense as
a statistical observable, $N_{LB} >> 1$.
For typical values of $\rho_{sol}=1$ gr/cm$^3$, $m_{sol}\sim20$ amu,
$\rho_{LB}=1$ (which correspond to water), and $h$ in the range $1-10^2$ nm,
 this yields $S_{N} \sim 10^{4}-10^{6}$.
This shows that the neglect of many-body fluctuations inherent to the 
single-particle Boltzmann representation is still justified even at the
nanoscopic scale of the lattice spacing.

We will focus here on the {\it fast} translocation regime, in
which the translocation time $t_X$ is much smaller than the
Zimm time, $t_Z$, i.e. the typical relaxation time of the 
polymer towards its native (minimum energy, maximum entropy) configuration. 
Under fast-translocation conditions, the many-body
aspects of the polymer dynamics cannot be ignored
because different beads along the chain do not move independently.
As a result, simple one-dimensional Brownian models do not apply 
\cite{KARDAR}. In addition to many-body solute-solute
interactions, the present approach also takes full account of
many-body solute-solvent hydrodynamic interactions. 
The conditions for fast-translocation regime can be 
appraised as follows. The translocation time is estimated 
by equating the driving force, $F_{pull}$, to
the drag force exerted by a solvent with dynamic viscosity 
$\eta$ on a polymer
with radius of gyration $R$, $F_{drag} \sim\frac{6 \pi \eta R^2}{t_X}$.
This yields $t_X \sim \frac{6 \pi \eta R^2}{F_{pull}}$. 
Since the Zimm time is given by $t_Z \sim \frac{0.4 \eta R^3}{k_B T}$, 
the fast-translocation condition $t_X \ll t_Z$ becomes:
\begin{equation}
\frac{F_{pull} R}{k_BT} \gg \frac{6 \pi}{0.4} \sim 50 
\end{equation}
Our reference simulation is performed with
$F_{pull}/m=0.02 \Delta x/\Delta t^2$ and $k_B T/m=10^{-4} \Delta x^2/\Delta t^2$, with $m$ the mass of one bead (monomer) of the polymer.
The polymer length is in the range $20 \leq N \leq 400$ beads. 
It can be readily checked that by assuming $R \sim N^{0.6}$ 
our set of parameters falls safely within the fast translocation regime.
However, for $k_B T/m=10^{-3}$, $F_{pull} R/k_B T$ is of the order of
$10^2-10^3$ which is much closer to breaking the above condition.

The main parameters of the simulation are (in LB units)
$\sigma=1.8 $ and $\epsilon= 10^{-4} $ for the Lennard-Jones potential.
The bond length among the beads is set at $r_0=1.2$. 
According to these values, the Lennard-Jones time-scale,
$\tau_{LJ}=\sigma/\sqrt {2 \epsilon/m}$, is 
of the order of $\sim 100 \Delta t$. 
Thus, by choosing $M=5$ as a time-gap factor, we obtain $dt
\sim \tau_{LJ}/500$, which is adequate for the resolution of
the polymer dynamics. The solvent is set at a density $\rho =1$, with
a kinematic viscosity $\nu=0.1 \Delta x^2/\Delta t$ and a damping
coefficient $\gamma=0.1/\Delta t$. The flexional rigidity $\kappa$
for the angular potential between beads will be $10^{-4}$/rad.
In order to resolve the structure of the solvent accurately on the 
atomistic scale \cite{HORBA}, we should use a higher resolution of 
at least 3-4 orders of magnitude. This means resolving the radial
structure of the pore, a task that can only be undertaken by resorting
to parallel computing. It is nonetheless hoped, and verified
{\it a posteriori}, that this artificial magnification does not affect
adversely the most
significant dynamical and statistical properties of the translocation
process, by which we mean that eventually, the time-scale
of the simulated process may not be the same as in the physical process
of interest, but the simulated dynamics is related to the
physical dynamics by a simple rescaling of the time variable.

\subsection{Translocation time}

The most immediate quantity of interest in the translocation process
is the dependence of the translocation time on the polymer
length. This is usually expressed by a scaling law of the form
\[
\tau_X (N) \equiv t_x/dt = N^\alpha
\]
where $t_0$ is a reference time-scale, formally the translocation
time of a single monomer, and $\alpha$ a scaling exponent measuring the
degree of competition ($\alpha >1$) / cooperation
($\alpha<1$) of the various monomers in the chain.

We first turn to the derivation of the scaling behavior of the
translocation process in the case where hydrodynamic interactions are 
included. In order to take 
into account the statistical nature of the phenomenon, simulations 
of a large number of translocation events ($100$ up to $1000$) 
for each polymer length were carried out.
The ensemble of simulations is generated by different realizations 
of the initial polymer configuration.
The duration histograms were constructed by cumulating all events for
various lengths. Overall, our results are quite similar to the
corresponding experimental data for DNA translocation 
through a nanopore \cite{NANO}, which we discuss in more
detail in the following section. 

\begin{figure}[ht]
\begin{center}
\includegraphics[width=1.0\textwidth]{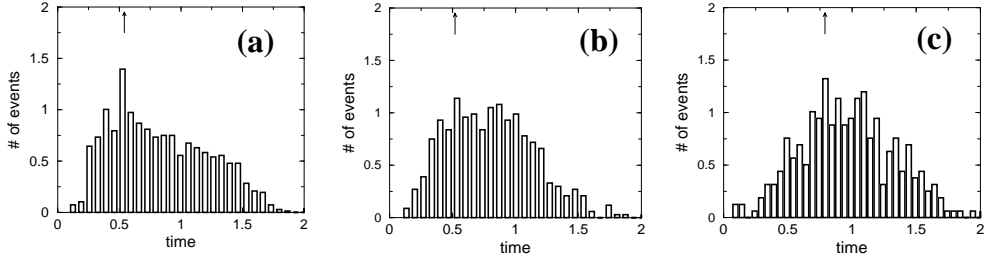}
\caption{\label{FIG4} Probability distributions of the translocation 
times for various lengths: (a) N =50, (b) N=100, and (c) N=300, 
respectively. Both axes are scaled to produce normalized probability
distributions. The arrows show the most probable translocation time
for each length.}
\end{center}
\end{figure}

At a next step, our data were shifted and scaled so that the
distribution curve starts at zero-time
and the total probability is equal to unity. 
The resulting distributions are on average not gaussians, but skewed 
towards long translocation times, consistent with experiment \cite{NANO}.
Therefore, the translocation time for each length is not assigned to the 
mean time, but to the most probable time, which is the position 
of the maximum in the histogram. In Fig.\ref{FIG4} the distribution of
all the events for polymer sizes N=50, 100 and 300 are shown. In 
this figure, the most probable translocation time for each length 
is denoted by an arrow.
From this analysis, a nonlinear relation between the most probable 
translocation time $\tau_0$ and the polymer length is obtained that
follows closely the theoretically expected scaling 
$\tau_X (N) \sim N^{\alpha}$, with $\alpha \sim 1.29$ (see Fig.\ref{FIG4}).

\begin{figure}[ht]
\begin{center}
\includegraphics[width=0.4\textwidth,angle=-90]{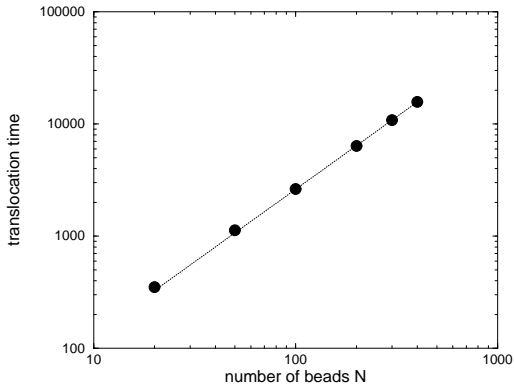}
\caption{\label{FIG5} Translocation time as a function of the number of 
monomers for the case with hydrodynamics. The straight line represents
 the power law with exponent $1.29$. Time is shown in units of the LB
timestep $\Delta t$.}
\end{center}
\end{figure}

\subsection{Dynamics with and without a solvent}

\begin{figure}
\begin{center}
\includegraphics[width=1.0\textwidth]{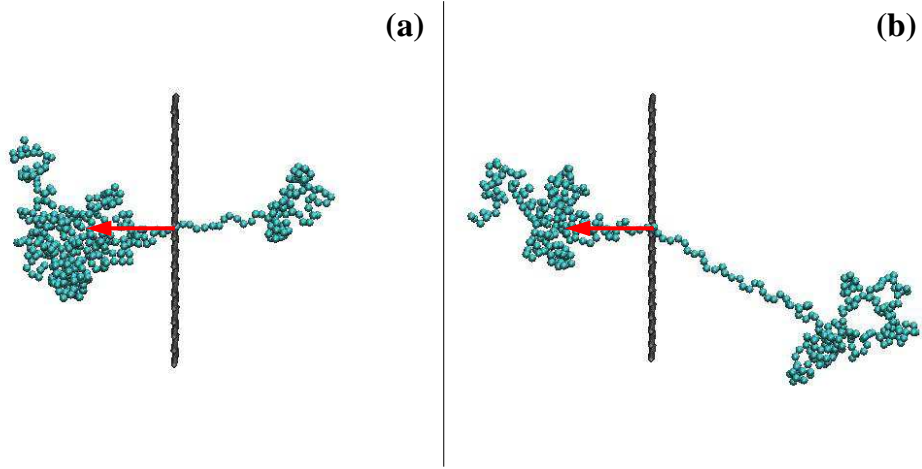}
\caption{\label{FIG6} Polymer configuration ($N=400$) corresponding to (a)
fast and (b) slow translocation events. Both snapshots are shown at a timestep
where the polymer (b) starts to slow down (see arrow in Fig.\ref{FIG7}).
$F_{pull}$ is applied at the hole region towards a direction indicated
by the arrow.}
\end{center}
\end{figure}

A closer inspection into the polymer dynamics reveals some interesting
features. The molecule shows a blob-like conformation on 
either side of the membrane as it moves
through the hole. It may either translocate very fast or move from 
one chamber to the other intermittently, with pauses.
Both types of events are present with and without a fluid solvent.
In addition, a careful analysis of all the translocated
chains unravels the difference between slower and faster translocation
within the same fast translocation regime. The nature of the variations 
in time is connected to the random fluctuations of the polymer throughout
its motion, rather than the temperature or its length.
These fluctuations are correlated to the entropic forces (gradient
of the free energy with respect to a conformational order parameter, 
typically the fraction of translocated beads, see $r(t)$ below) 
acting on both translocated and untranslocated parts of the polymer. 
In fact, when a solvent is present, the interplay between these forces 
and $F_{drag}$, $F_{pull}$ determines 
the motion and the shape of the chain and thereby the translocation time.
At some point part of the chain shapes up in an almost linear
conformation increasing in this way the entropic force acting on it. 
This eventually leads to deceleration of the whole chain. 
Fig.\ref{FIG6} shows an illustration of this argument, where a
polymer chain, surrounded by a solvent is represented at a 
time where it starts to slow down. 
In this figure, a polymer with the same length but different
initial configuration is also shown at the same time. 

\begin{figure}
 \begin{center}
\includegraphics[width=0.5\textwidth,angle=270]{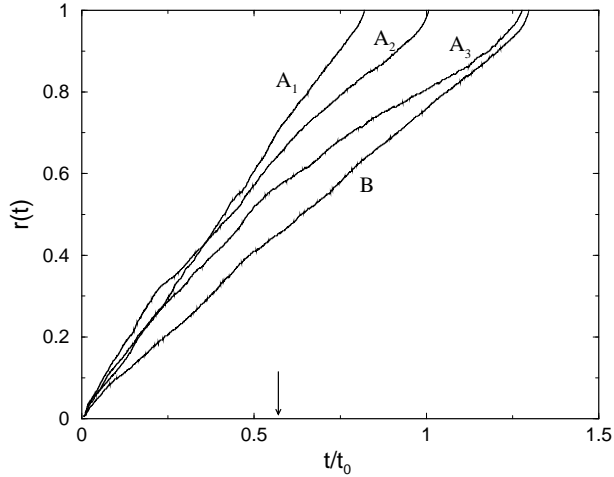}
 \caption{\label{FIG7}Progress in time of the number of translocated beads
for chains with $N=400$ monomers.
 Curves $A_1$, $A_3$ correspond to slow and fast translocation events
(polymers shown in Fig.\ref{FIG5}), while $A_2$ to an event related to the
most probable time. The initial configuration for the polymer
in the event $B$ is the same as for $A_2$, but in that case no hydrodynamic
 interactions are included. Time is scaled
with respect to the value of $\tau_0$ in the case with hydrodynamics.
The arrow indicates the timestep at which the snapshots in Fig.\ref{FIG6}
are shown.}
 \end{center}
 \end{figure}

It is very instructive to monitor the progress in time of the
number of translocated monomers $N(t)$. Note that $r(t) \equiv N(t)/N$
serves as a reaction-coordinate, with the translocation time 
defined by the condition $r(t_X)=1$. The translocated monomers for
processes with and without hydrodynamics are shown in Fig.\ref{FIG7}.
For the former, events related to the polymers of Fig.\ref{FIG6}
are shown (curves $A_1$, $A_3$), as well as that related to the
most probable time ($A_2$). The arrow in this figure 
indicates the timestep corresponding to the snapshots in Fig.\ref{FIG6}.
The translocation for a given polymer proceeds along a curve
virtually related to its initial configuration and its interactions with
the fluid. It is clearly visible that there is no general trend. 
The non-hydrodynamic case is in principle different, especially in terms 
of the time range which is larger.
This reveals the importance of hydrodynamic coherence.

Additional insight into the dynamics is obtained by altering the parameter
set. This has not yet been extensively explored, but it was found that
a choice of $k_BT=10^{-5}$, $F_{pull}=0.01$, and
$\epsilon=0.002$ leads to the frequent retraction of the polymer.
In other words, after having translocated a large fraction 
of its length, the polymer occasionally reverses its motion and 
anti-translocates away from the hole, never to find its way back into it. 
Moreover, we find that a polymer that retracted in the presence of
a solvent, manages to fully translocate if the solvent is absent.
It is interesting to observe that, in principle, no such type of
anti-translocating behavior has been observed for short polymers.
This indicates that hydrodynamics significantly speed-up 
and alter the nature of
translocation, especially for long polymers at low temperatures.
This highly irregular dynamics escapes any scaling or
statistical analysis, as well as dynamic Monte Carlo simulations
 \cite{CHERN}, and can only be revealed by self-consistent many-body 
hydro-dynamic simulations.


\section{DNA translocation through a nanopore}
\label{relate2dna}

The translocation of biopolymers, such as DNA and RNA plays
a major role in many important biological processes, such as
viral infection by phages, inter-bacterial DNA transduction, and
gene therapy \cite{TRANSL}. The importance of this process has 
spawned a number of {\it in vitro} experiments, aimed
at exploring the translocation process through micro-fabricated channels
under the effects of an external electric field, or through protein
channels across cellular membranes \cite{EXPRM}.
In particular, recent experimental work has focused on the possibility
of fast DNA-sequencing by reading the base sequence as the polymer 
passes through a nanopore.
Some universal features of DNA translocation can be analyzed by
means of suitably simplified statistical schemes \cite{statisTrans}
and non-hydrodynamic coarse-grained or microscopic models
\cite{DynamPRL,Nelson}. However, a quantitative description 
of this complex phenomenon calls for state-of-the art modeling of the 
type described above. Accordingly, we explore here to what extent the
results discussed above for the generic situation of polymer translocation
apply to the DNA case.

First, we note that, as already mentioned in the
previous section, our results are quite similar to the
experimental data for DNA translocation through a nanopore \cite{NANO}.
Three different interpretations of the current model are physically 
plausible:\\
(a) Following the framework used in recent studies of DNA packing in
bacteriophages \cite{beadParam1}, one monomer in our simulation can 
be thought of as representing 
a DNA segment of about 8 base-pairs, that is, each bead has a diameter 
of 2.5 nm, the hydrated diameter of B-DNA in physiological
conditions.  \\
(b) It is also physically plausible to assume that a bead 
represents a portion of DNA equivalent to  
its persistence length of about 50 nm, which translates into  
mapping one bead to $\sim$150 base-pairs. \\
(c) Alternatively, 
as is typically done in 
simulations of the $\lambda$-phage DNA in solution \cite{beadParam2}, 
one bead can be taken to correspond to  
$\sim 10^3$ base-pairs.\\
In all three cases $h = \Delta x$ is equal to the bead size, while
the pore, having a width of 2$\Delta x$, will be different from 
the pores used experimentally, either smaller or larger. In addition,
 the coarse graining model that handles the DNA molecules indicates
that the MD timescale is stretched over the physical process. 
A direct comparison between our probability distributions for 
polymer translocation and the experimental results sets a different
MD timestep for the cases (a), (b), and (c) which is of the order of 
3 nsec, 100 nsec and 5 $\mu$sec, respectively, leading to a LB
 timestep $\Delta t=5 dt$ of 15 nsec, 500 nsec, and 25 $\mu$sec.
It is difficult at this stage to assign a unique interpretation of our model 
in relation to a physical system.  A thorough exploration of the 
parameter space is required before such an assignment can be made 
convincingly. This is beyond the scope of the present work but will 
be reported in future publications.

\begin{figure}
\begin{center}
\includegraphics[width=1.0\textwidth]{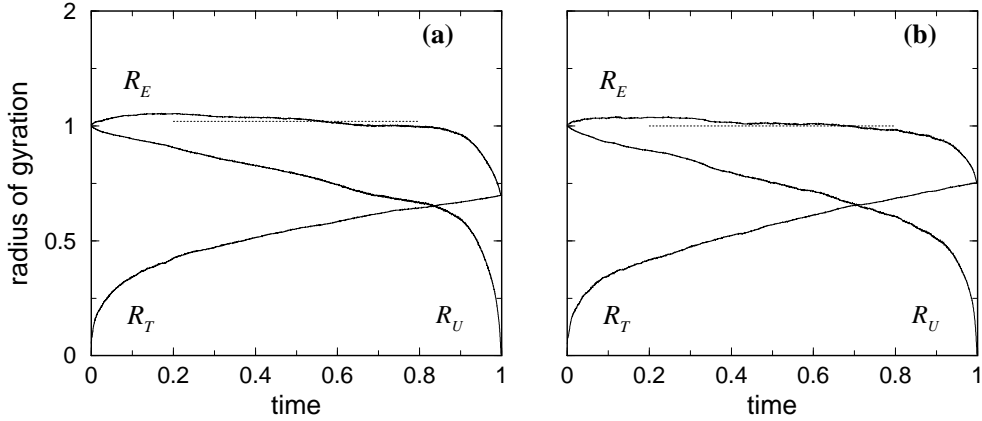}
\caption{\label{FIG8} Translocated ($R_{T}$), untranslocated ($R_{U}$) 
and effective ($R_{E}$) radii of gyration for 
the (a) non-hydrodynamic and (b) hydrodynamic cases with
 $N=400$. All radii are normalized with respect to the initial value
$R_{U}(t=0)$. Time is also scaled with respect to the total translocation
time $t_X$ for each of the events (a) and (b).
The dotted lines denote the regions where $R_{E}$ is nearly
constant (see text).
}
\end{center}
\end{figure}

A second encouraging comparison is that the scaling we found 
for the translocation time with polymer length 
(with exponent $\alpha = 1.29$) is quite close 
to the experimental measurement for DNA translocation \cite{NANO}
($\alpha = 1.27 \pm 0.03$). 
Beyond the apparent consistency between experiment and theory, 
additional insight can be gained by analyzing the polymer 
dynamics during translocation. 

A hydrodynamic picture of DNA translocation has been presented 
in ref. \cite{NANO}. In this work, the authors assume that the electric 
field drive is in balance with the Stokes drag exerted by the solvent 
on the blob configuration of the polymer, that is
$$F_{pull} = 6\pi\rho \nu  {d R_U^2\over \tau}$$ 
where $\rho$ is the density, $\nu$ the kinematic viscocity, $\tau$ the 
translocation time, and
$R_U$ the translocated part of the radius of gyration.
In order for this balance to apply at all times, it is clear that
$R^2$ must be constant in time, hence it cannot be identified neither 
with the translocated nor with the untraslocated gyration radius of DNA.
To this end, in Fig.\ref{FIG8} we represent the time-evolution
of the radii of gyration for the two sections of DNA, 
$R_U(t)$ and $R_T(t)$, the untranslocated ($U$) 
[with $x>h N_x/2$] and translocated $(T)$ [with $x<h N_x/2$] parts, 
respectively.
In order to identify a time-invariant radius, we define the
$R_I (t)=QN_{I}^{\zeta}(t)$, where $I=U,T$ stands for the 
untranslocated and translocated parts and $N_{I}(t)$
is the corresponding number of monomers. The exponent
$\zeta\sim0.6$ is the same as previously noted and $Q$ is a constant 
(for long enough polymers).
If translocation could be described by the dynamics of a single-blob object, 
characterized by an effective radius of gyration, defined as 
$$R_{E}(t)= c N^{\zeta}(t),$$ then this quantity should be constant 
in time. Since the $N=N_{T}(t)+N_{U}(t)$ holds for all $t$, 
the above relations lead to 
\begin{equation}
R_{E}(t)=[R_T^{1/\zeta} (t)+R_U^{1/\zeta}(t)]^{\zeta}  = const
\label{R_total}
\end{equation}
We focus first on the case without hydrodynamics, Fig.\ref{FIG8}(a).
For very small chains $R_I(t)$ does not scale as
$N_{I}^{0.6}$ and the definition for $R_{E}$
is not valid at the first and last $\sim15-20\%$ 
parts of the event, during which the untranslocated
and the translocated parts, respectively, are small.
Outside these limits, 
$R_{E}(t)$ as obtained from the definition (\ref{R_total}), with the
values of $R_U(t), R_T(t)$ directly taken from the simulations, is 
indeed approximately constant.
In addition, the values $R_T(t=t_X)$ and $R_U(t=0)$ do not coincide,
since the former is lower than the latter. 
This is also the case when a solvent is present, Fig.\ref{FIG8}(b). 
Thus, regardless of the dynamic pathway and the
different conformations the chain may possess during
translocation, once the event is completed the polymer is
more compact than at $t=0$.
Comparison of the cases with and without hydrodynamics reveals
that in the latter case the polymer becomes up to $\sim7\%$
more confined than when a solvent is added.
The untranslocated part of the radius of gyration at the 
end of the process shows an abrupt drop. As a consequence
the polymer $t=t_X$ does not fully recover its initial volume.
It it plausible, that by allowing the polymer to further advance
in time, $R_T(t=t_X)$ will become similar to
$R_U(t=t_0)$, but this remains to be examined.
Nevertheless, in this work we have been interested mainly on the 
chain dynamics related to the first passage times, 
which correspond to the exact period of time needed until all
the beads have translocated.

\section{Conclusions}
\label{outlook}

We have presented a multiscale methodology based on the
concurrent coupling of constrained molecular dynamics for the solute
biopolymers with a lattice Boltzmann treatment of solvent dynamics.
Owing to the dual field-particle nature of the Lattice Boltzmann 
technique, this coupling proceeds seamlessy in time and only requires 
standard interpolation/extrapolation for information-transfer in
physical space. This multiscale methodology has been applied to the 
case of polymer translocation through a nanopore, with special emphasis 
on the role of hydrodynamic coherence on the dynamic and statistical 
properties of the translocation process. It is found that hydrodynamic 
interactions play a major role in accelerating the translocation process, 
especially for long molecules at low temperature. 

An attempt to connect these results to the process of DNA 
translocation through a nanopore revealed certain similarities with experiment,
especially in the scaling law of the translocation time with polymer length.
The presence of hydrodynamic interactions lead to a decrease in 
the translocation times, compared to the cases without a fluid solvent. 
Inspection of the variation of the translocated beads and the
radii of gyration with time reveals interesting aspects 
of the DNA dynamics during translocation.

Future directions for the simulations include the
detailed study of the effects of temperature, finite-length and geometrical 
details of the nanopore geometry, as well as electrostatic interactions of the 
DNA molecule with the surrounding fluid.
To this end, resort to parallel computing is mandatory, and we expect
the favourable properties of LB towards parallel implementations to
greatly facilitate the task. Work along these lines is currently in
progress.

\vspace{0.7cm}

{\bf Acknowledgements}
The authors wish to thank C. Pierleoni for valuable discussions
and help with the validation tests.

\end{document}